\documentclass[aps,pre,showpacs,twocolumn]{revtex4}
\usepackage{bm}
\usepackage{graphicx}
\bibstyle{approve.bib}
\usepackage{amssymb}
\usepackage{amsmath}
\usepackage{esint}
\usepackage{epstopdf}
\usepackage{color}
\usepackage{gensymb}
\usepackage{mathtools}
\usepackage{suffix}

\newcommand{\be}{\begin{equation}}
\newcommand{\ee}{\end{equation}}
\newcommand{\bea}{\begin{eqnarray}}
\newcommand{\eea}{\end{eqnarray}}

\newcommand{\br}{\mathbf{r}}

\newcommand{\e}{\varepsilon}

\newcommand{\tv}{\tilde{v}}

\begin{document}

\title{Multivalent cation induced attraction of anionic polymers by like-charged pores}

\author{Sahin Buyukdagli$^{1}$\footnote{email:~\texttt{Buyukdagli@fen.bilkent.edu.tr}}  and 
T. Ala-Nissila$^{2,3}$\footnote{email:~\texttt{Tapio.Ala-Nissila@aalto.fi}}}
\affiliation{$^{1}$Department of Physics, Bilkent University, Ankara 06800, Turkey\\
$^{2}$Department of Applied Physics and COMP Center of Excellence, Aalto University School of Science, 
P.O. Box 11000, FI-00076 Aalto, Espoo, Finland\\
$^{3}$Departments of Mathematical Sciences and Physics, Loughborough University, Loughborough, 
Leicestershire LE11 3TU, United Kingdom}


\begin{abstract}
The efficiency of nanopore-based polymer sensing devices depends on the fast capture of anionic polyelectrolytes by negatively charged pores. This requires the cancellation of the electrostatic barrier associated with repulsive polymer-pore interactions. We develop a correlation-corrected theory to show that the barrier experienced by the polymer can be efficiently overcome by the addition of multivalent cations into the electrolyte solution. Cation adsorption into the pore enhances the screening ability of the pore medium with respect to the bulk reservoir which translates into an attractive force on the polymer. Beyond a critical multivalent cation concentration, this correlation-induced attraction overcomes the electrostatic barrier and triggers the adsorption of the polymer by the like-charged pore. It is shown that like-charge polymer-pore attraction is suppressed by monovalent salt but enhanced by the membrane charge strength and the pore confinement. Our predictions may provide enhanced control over polymer motion in translocation experiments. 

\end{abstract}


\date{July 1, 2017}
\maketitle

\section{Introduction}

The interaction of charged solutes with membrane nanopores plays a central role in biological processes and the functioning of biosensing methods~\cite{Tapsarev}. Among these techniques, drift-driven polymer translocation through biological and synthetic nanopores has been a central focus for over the past two decades~\cite{e1,e2,e3,e4,e5}. This approach consists of reading the polymer sequence through the ionic current alterations induced by the translocating polyelectrolyte. The precision of the method requires an accurate control over the polymer dynamics governed by entropic and electrohydrodynamic polymer-pore and polymer-liquid interactions. The characterization of these interactions is thus of major importance for the optimization of polymer translocation based sequencing devices.

The electrohydrodynamics of polymer-liquid interactions and entropic effects associated with conformational polymer fluctuations have been scrutinized by simulations~\cite{sim1,sim2,sim3,sim4,sim5} and theoretical models~\cite{the1,the2,the3,the4,the5,the6,the7}. However, the direct electrostatic coupling between the polymer and the membrane nanopore has been mostly overlooked. This is a strong theoretical limitation; the majority of translocation experiments involve negatively charged polymers driven through anionic silicon based membrane nanopores~\cite{e6,e7,e8,e9,e10}. Thus, the like-charge polymer-pore interactions are expected to induce a barrier that may severely limit polymer capture by the pore. This was indeed explicitly shown by our recent mean-field (MF) polymer translocation model~\cite{mf}. At this point, it should be noted that the presence of the electrostatic barrier hinders the optimal functioning of the  polymer translocation method whose efficiency requires the fast capture of the polymer from the reservoir. Thus, the optimization of this sequencing technique necessitates the removal of the electrostatic barrier induced by direct like-charge polymer-pore interactions.  

In this article, we show that the electrostatic barrier experienced by the polymer can be efiiciently overcome 
by adding multivalent counterions into the solution. The counterion attraction by the anionic pore walls results in a cationic excess in the pore. Due to this ionic abundance, the pore electrolyte can screen the polymer charges more efficiently than the reservoir solution. This lowers the polymer's free energy in the pore with respect to the reservoir medium and translates into an attractive force. Beyond a critical concentration of multivalent cations, this correlation-induced force takes over the repulsive barrier and triggers an electrostatic attraction on the polymer by the like-charged pore.

The main novelty in the present work concerns the fact that treating the electrostatics from multivalent ions requires the formulation of polymer-pore interactions beyond the MF-Poisson-Boltzmann level. To this end, we make use of a test-charge approach that was introduced in Ref.~\cite{Buyuk2017} for general geometry. In Section~\ref{testch}, we express the characteristic equations of the test-charge theory in the specific geometry of the polymer-pore complex. The polymer grand potential characterizing electrostatic polymer-pore interactions is composed of the MF-level interaction term and the polymer self-energy bringing one-loop-level charge correlations. The MF component is calculated within an improved Donnan approximation in Section~\ref{mfgrand}. In the computation of the polymer self-energy, the main technical complication arises from the cylindrical geometry of the system where the one-loop-level kernel equation satisfied by the electrostatic propagator cannot be solved analytically. In order to overcome this difficulty, we develop an analytical Wentzel-Kramers-Brillouin (WKB) solution scheme explained in Section~\ref{self} in detail. Within this beyond-MF theory, in Section~\ref{res}, we throughly investigate electrostatic correlation effects on polymer-pore interactions. We summarize our results and discuss potential improvements to our theory in Conclusions.

\section{Theory}
\label{grpotl}

\begin{figure}
\includegraphics[width=1.\linewidth]{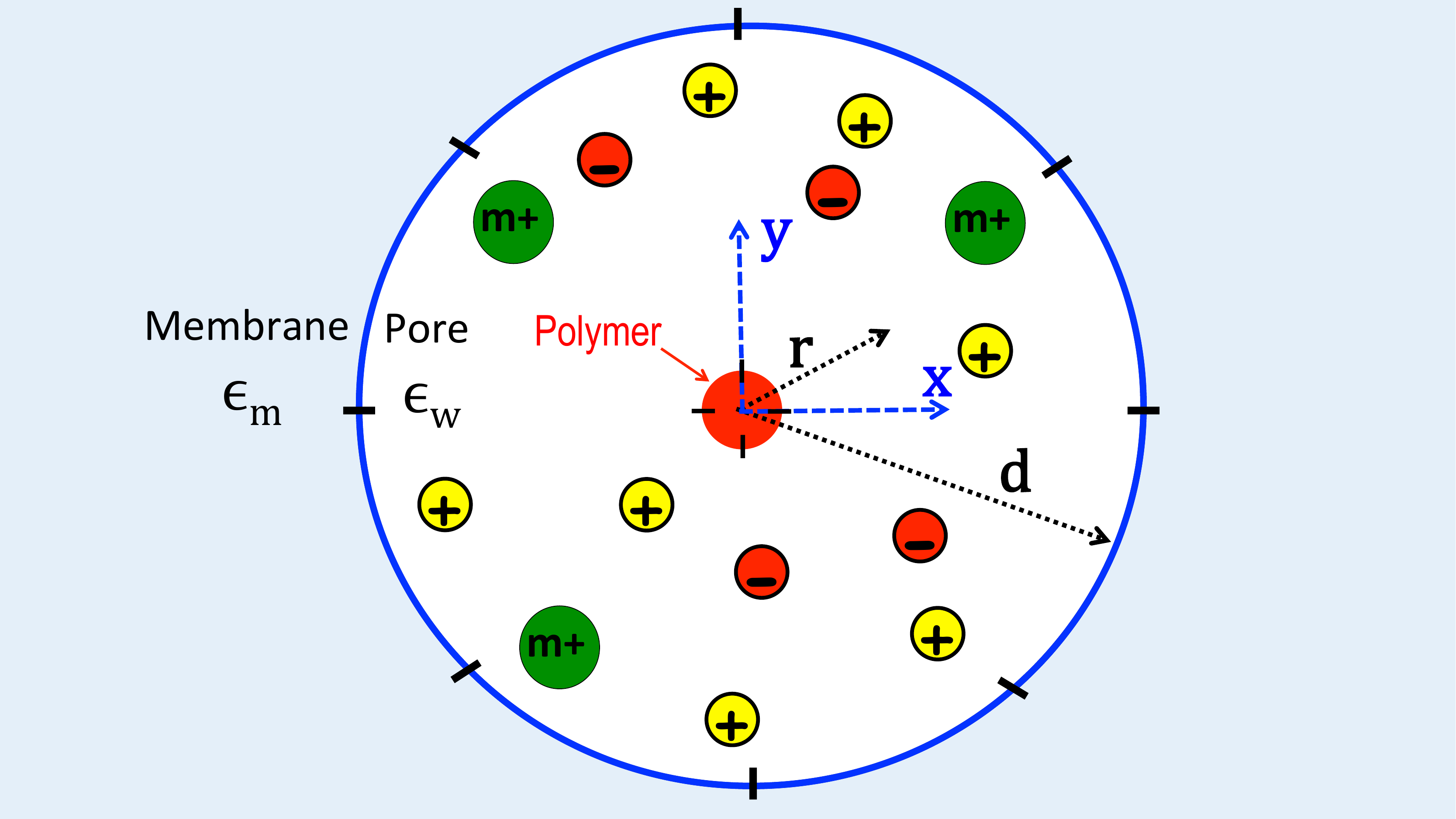}
\caption{(Color online) Schematic representation of the polyelectrolyte with line charge density $\tau$ located on the axis of the cylindrical nanopore. The pore has radius $d$ and fixed negative surface charge density $-\sigma_{\rm m}$.  The polymer portion inside the pore has length $l_{\rm p}$. The membrane and pore dielectric permittivities are respectively $\e_{\rm m}=2$ and $\e_{\rm w}=80$.}
\label{fig1}
\end{figure}

In this section, we introduce a beyond-MF electrostatic theory of polymer-pore interactions in mixed electrolytes. To this end, we calculate the polymer grand potential that determines the electrostatic cost for the capture of the polymer by the nanopore. Figure \ref{fig1} 
displays the charge composition of the system. The cylindrical nanopore of radius $d$ and negative wall charge density $-\sigma_{\rm m}$ is connected to a bulk ion reservoir. The pore and the reservoir contain a mixed electrolyte solution. The  solution is composed of $p$ ionic species. The species $i$ has valency $q_i$ and reservoir concentration $\rho_{{\rm b} i}$. For the sake of simplicity, we consider the polymer as a line charge with density $\tau=2\pi a\sigma_{\rm p}$ located along 
the pore axis. Here, $a=1$ nm and $\sigma_{\rm p}=0.4$ $\mbox{e/nm}^2$ correspond respectively to the radius and surface charge density of the corresponding  cylindrical double-stranded (ds) DNA molecule~\cite{not2}.  

The calculation of the polymer grand potential will be based on the test-charge approach previously developed for general geometry in Ref.~\cite{Buyuk2017}. In Section~\ref{testch}, we briefly review the test-charge theory and recast the characteristic equations of state in the cylindrical geometry associated with the polymer-pore complex. The polymer grand potential includes a repulsive MF term accounting for the direct interaction between the polymer and pore charges, and the polymer self-energy that brings charge correlation effects. The MF and self-energy components are derived respectively in Sections~\ref{mfgrand} and~\ref{self}.

\subsection{Electrostatic theory of polymer-pore interactions}
\label{testch}

Here we briefly review the test-charge approach of Ref.~\cite{Buyuk2017} and express the polymer grand potential in the geometry corresponding to
Fig. 1. In the following calculation, we approximate the nanopore as an infinitely long cylinder in the $z$ direction. 
According to the test-charge theory, the polymer grand potential is composed of two components, namely
\be
\label{1}
\Delta\Omega_{\rm p}=\Omega_{\rm MF}+\Delta\Omega_{\rm s}.
\ee
The first term of Eq.~(\ref{1}) is the MF component associated with the direct coupling between the polymer and pore charges. 
Rescaled by the thermal energy, this term reads
\be
\label{2}
\beta\Omega_{\rm MF}=\int\mathrm{d}\br\sigma_{\rm p}(\br)\phi_{\rm m}(\br).
\ee
In Eq.~(\ref{2}), the charge density function of the polymer is
\be\label{3}
\sigma_{\rm p}(\br)=-\frac{\tau}{r}\delta(r-r_{\rm p})\delta(\varphi-\varphi_{\rm p})\theta(z)\theta(l_{\rm p}-z),
\ee
where $r_{\rm p}$ stands for the radial distance of the polymer from the pore axis and the polar angle $\varphi_{\rm p}$ 
indicates its the location on the $xy$ plane. Thus, for the time being, we do not restrict the polymer position to the pore axis but 
simply assume that the polymer is oriented parallel with the $z$ axis. In Eq.~(\ref{3}), $\phi_{\rm m}(\br)$ is the average 
potential induced exclusively by the fixed charges on the membrane wall. Thus, this potential solves the PB equation 
\be\label{4}
\nabla\e(r)\nabla\phi_{\rm m}(\br)+\frac{e^2}{k_{\rm B}T}\sum_{i=1}^pq_in_i(\br)=-\frac{e^2}{k_{\rm B}T}\sigma_{\rm m}(r),
\ee
where we introduced the dielectric permittivity profile
\be\label{5}
\e(r)=\e_{\rm w}\theta(d-r)+\e_{\rm m}\theta(r-d),
\ee
with the membrane permittivity $\e_{\rm m}=2$ and the pore permittivity $\e_{\rm w}=78$. In Eq.~(\ref{4}), 
$e$ is the electron charge, $k_{\rm B}$ the Boltzmann constant, and $T=300$ K the solvent temperature. Furthermore, the function
\be\label{6}
n_i(\br)=\rho_{{\rm b}i}e^{-q_i\phi_{\rm m}(\br)}\theta(d-r)
\ee
corresponds to the ion number density distribution in the pore, with $\theta(x)$ the Heaviside step function. 
Finally, in Eq.~(\ref{4}), the density distribution of the fixed charges on the membrane wall reads
\be\label{7}
\sigma_{\rm m}(r)=-\sigma_{\rm m}\delta(r-d),
\ee
We note that in the bulk reservoir where the average potential vanishes {\it i.e.} $\phi_{\rm m}(\br)=0$, the 
MF grand potential of Eq.~(\ref{2}) vanishes as well, {\it i.e.} $\Omega_{\rm MF}=0$.

The second term in Eq.~(\ref{1}) corresponds to the difference between the self-energy of the polymer located in the pore and the bulk reservoir. This self-energy rescaled with the thermal energy reads
\be
\label{8}
\beta\Delta\Omega_{\rm s}=\frac{1}{2}\int\mathrm{d}\br\mathrm{d}\br'\sigma_{\rm p}(\br)\left[v(\br,\br')-v_{\rm b}(\br-\br')\right]\sigma_{\rm p}(\br'),
\ee
where the electrostatic propagator $v(\br,\br')$ solves the one-loop level kernel equation
\be\label{9}
\nabla\e(r)\nabla v(\br,\br')-\frac{e^2}{k_{\rm B}T}\sum_{i=1}^pq_i^2n_i(\br)v(\br,\br')=-\frac{e^2}{k_{\rm B}T}\delta(\br-\br').
\ee
In Eq.~(\ref{8}), we used the electrostatic propagator in the bulk. This corresponds to the spherically symmetric DH potential 
$v_{\rm b}(\br-\br')=\ell_{\rm B}e^{-\kappa_{\rm b}|\br-\br'|}/|\br-\br'|$, with the Bjerrum length 
$\ell_{\rm B}\approx7$ {\AA} and the DH screening parameter
\be\label{10}
\kappa_{\rm b}^2=4\pi\ell_{\rm B}\sum_{i=1}^p\rho_{{\rm b}i}q_i^2.
\ee
We also note that Eqs.~(\ref{4}) and~(\ref{9}) should be solved with the electroneutrality condition in the reservoir, given by
\be
\label{neut}
\sum_{i=1}^p\rho_{{\rm b}i}q_i=0.
\ee

Due to the cylindrical symmetry of Eqs.~(\ref{5}) - (\ref{7}), the electrostatic potential depends solely on the radial distance 
$r$, i.e. $\phi_{\rm m}(\br)=\phi_{\rm m}(r)$. Moreover, within the same symmetry, the electrostatic Green's function can be 
Fourier expanded as
\be\label{11}
v(\br,\br')=\sum_{n=-\infty}^{+\infty}e^{in(\varphi-\varphi')}\int_{-\infty}^{\infty}\frac{\mathrm{d}k}{4\pi^2}e^{ik(z-z')}\tv_n(r,r';k).
\ee
Evaluating the integrals in Eqs.~(\ref{2}) and~(\ref{8}) with Eqs.~(\ref{3}) and~(\ref{11}), the grand potential components simplify to
\bea\label{12}
\beta\Omega_{\rm MF}(r_{\rm p},l_{\rm p})&=&-l_{\rm p}\tau\phi_{\rm m}(r_p);\\
\label{13}
\beta\Delta\Omega_{\rm s}(r_{\rm p},l_{\rm p})&=&\frac{l_{\rm p}\tau^2}{4\pi}\sum_{n=-\infty}^{+\infty}
\int_{-\infty}^{\infty}\mathrm{d}k\frac{2\sin^2(kl_{\rm p}/2)}{\pi k^2l_{\rm p}}\\
&&\hspace{1.cm}\times\left[\tv_n(r_{\rm p},r_{\rm p};k)-\tv_{{\rm b},n}(r_{\rm p},r_{\rm p};k)\right]\nonumber.
\eea
Moreover, the PB Eq.~(\ref{4}) and the kernel Eq.~(\ref{9}) take the radial form
\bea\label{14}
&&\frac{k_{\rm B}T}{e^2}\frac{1}{r}\partial_r\left[r\e(r)\partial_r\phi_{\rm m}(r)\right]+\sum_{i=1}^pq_in_i(r)=\sigma_{\rm m}\delta(r-d);\nonumber\\
&&\\
\label{15}
&&\left\{\frac{1}{r}\partial_rr\e(r)\partial_r-\e(r)\left[\frac{n^2}{r^2}+k^2+\kappa^2(r)\right]\right\}\tv_{n}(r,r';k)\nonumber\\
&&=-\frac{e^2}{k_{\rm B}T}\frac{1}{r}\delta(r-r'),
\eea
with the local screening function
\be
\label{17}
\kappa^2(r)=4\pi\ell_{\rm B}\sum_{i=1}^pq^2_in_i(r).
\ee
The boundary conditions associated with the PB Eq.~(\ref{14}) are 
Gauss' law at the pore wall, and the vanishing electric field condition in the mid-pore,
\be
\label{18}
\phi'_{\rm m}(d^-)=-4\pi\ell_{\rm B}\sigma_{\rm m}\;;\hspace{5mm}\phi'(0)=0.
\ee
Finally, the matching conditions to be satisfied by the solution of the kernel Eq.~(\ref{15}) read
\bea\label{19}
&&\lim_{r\to d^+}\tv_{n}(r,r';k)=\lim_{r\to d^-}\tv_{n}(r,r';k);\\
\label{20}
&&\lim_{r\to r'^+}\tv_{n}(r,r';k)=\lim_{r\to r'^-}\tv_{n}(r,r';k);\\
\label{21}
&&\lim_{r\to d^+}\e(r)\partial_r\tv_{n}(r,r';k)=\lim_{r\to d^-}\e(r)\partial_r\tv_{n}(r,r';k);\\
\label{22}
&&\lim_{r\to r'^+}\partial_r\tv_{n}(r,r';k)-\lim_{r\to r'^-}\partial_r\tv_{n}(r,r';k)=-\frac{4\pi\ell_{\rm B}}{r'}.\nonumber\\
\eea

In order to evaluate the polymer grand potential components in Eqs.~(\ref{12}) and~(\ref{13}), we have to calculate the average 
potential $\phi_{\rm m}(r)$ solving Eq.~(\ref{14}) and the electrostatic propagator $\tv_{n}(r,r';k)$ solution to Eq.~(\ref{15}). 
We do not have exact analytic solutions to Eqs.~(\ref{14}) and~(\ref{15}). Below, we explain the analytical solution of these 
electrostatic equations within the Donnan and WKB approximations.

\subsection{Computing the mean field grand potential $\Omega_{\rm MF}(r_{\rm p},l_{\rm p})$ within Donnan approximation}
\label{mfgrand}

In order to compute the MF component in Eq.~(\ref{12}),  we will solve the PB Eq.~(\ref{14}) within an improved 
Donnan approximation. At the first step, in Eq.~(\ref{14}), we set $\phi_{\rm m}(r)=\phi_{\rm D}$,
where $\phi_{\rm D}$ is the constant Donnan potential, and integrate the resulting equation over the cross-section of the pore. 
This leaves us with the relation
\be\label{23}
\sum_{i=1}^p\rho_{{\rm b}i}q_ie^{-q_i\phi_{\rm D}}=\frac{2\sigma_{\rm m}}{d},
\ee
whose solution yields the Donnan potential $\phi_{\rm D}$. At the next step, we improve the Donnan approximation by accounting for the potential variations in the pore. We express the average potential as
\be\label{24}
\phi_{\rm m}(r)=\phi_{\rm D}+\delta\phi(r),
\ee
inject Eq.~(\ref{24}) into the PB Eq.~(\ref{14}), and Taylor expand the latter in terms of the correction term $\delta\phi(r)$. 
Using Eq.~(\ref{23}) and defining the Donnan screening parameter
\be\label{25}
\kappa_{\rm D}^2=4\pi\ell_{\rm B}\sum_{i=1}^p\rho_{{\rm b}i}q_i^2e^{-q_i\phi_{\rm D}},
\ee
one gets the differential equation $\left(r^{-1}\partial_rr\partial_r-\kappa_{\rm D}^2\right)\delta\phi(r)
=-8\pi\ell_{\rm B}\sigma_{\rm m}/d$. Imposing the boundary conditions in Eq.~(\ref{18}), the solution to this differential equation reads
\be
\label{26}
\delta\phi(r)=\frac{4\pi\ell_{\rm B}\sigma_{\rm m}}{\kappa_{\rm D}}
\left[\frac{2}{\kappa_{\rm D}d}-\frac{\mathrm{I}_0(\kappa_{\rm D}r)}{\mathrm{I}_1(\kappa_{\rm D}d)}\right],
\ee
where $I_n(x)$ is the modified Bessel function of the first kind~\cite{math}. Inserting Eq.~(\ref{24}) together with Eq.~(\ref{26}) into the MF grand potential~(\ref{12}), the latter takes the form
\bea\label{27}
\beta\Omega_{\rm MF}(r_{\rm p},l_{\rm p})&=&-l_{\rm p}\tau\phi_{\rm D}\\
&&-l_{\rm p}\tau\frac{4\pi\ell_{\rm B}\sigma_{\rm m}}{\kappa_{\rm D}}\left[\frac{2}{\kappa_{\rm D}d}-\frac{\mathrm{I}_0(\kappa_{\rm D}r_{\rm p})}{\mathrm{I}_1(\kappa_{\rm D}d)}\right].\nonumber
\eea

In Ref.~\cite{mf}, the MF grand potential in Eq.~(\ref{27}) was computed within the same approach for symmetric monovalent electrolytes and the accuracy of the improved Donnan approximation was shown by comparison with the exact solution of the PB Eq.~(\ref{14}). At this point, we note that due to the negative sign of the membrane charges, the pore potential of Eq.~(\ref{24}) is negative. Thus, the MF grand potential is positive and its magnitude rises steadily with the polymer length $l_{\rm p}$ in the pore. This behaviour accounts for the MF level electrostatic barrier experienced by the polymer during its penetration into the pore. We calculate next the self-energy component in Eq.~(\ref{13}) that brings charge correlations into the MF interaction picture.

\subsection{Computing the polymer self-energy $\Delta\Omega_{\rm s}(r_{\rm p},l_{\rm p})$ within WKB approximation}
\label{self}

Here, we compute the self-energy component of Eq.~(\ref{13}) of the polymer grand potential in Eq.~(\ref{1}). This requires the solution of the kernel Eq.~(\ref{15}). The homogeneous solutions to this equation can be in principle computed numerically. However, due to high memory requirements, the numerical scheme explained in Appendix~\ref{apx1} cannot be used for the calculation of the polymer grand potential at finite polymer length $l_{\rm p}$. Thus, in Section~\ref{hom}, the homogeneous solutions to Eq.~(\ref{15}) are derived within the WKB approach. In Section~\ref{part}, in terms of these homogeneous solutions, we calculate the particular solution to Eq.~(\ref{15}) that satisfies the boundary conditions of Eqs.~(\ref{19})-(\ref{22}). Finally in Section~\ref{polsel}, this particular solution is used for the computation of the polymer self-energy in Eq.~(\ref{13}).

\subsubsection{Homogeneous solution of the kernel Eq.~(\ref{15})}
\label{hom}

In order to solve the radial kernel Eq.~(\ref{15}), we have to find first the homogeneous solutions to the equation
\be
\label{28}
\left\{\frac{1}{r}\partial_rr\partial_r-\left[\frac{n^2}{r^2}+k^2+\kappa^2(r)\right]\right\}\tv_n(r,r';k)=0.
\ee
We note that the local screening function $\kappa(r)$ appearing in Eq.~(\ref{28}) will be calculated with the 
potential in Eq.~(\ref{24}) of the improved Donnan approximation. In the ion-free membrane region located at $r>d$, one has $\kappa(r)=0$. Therefore, inside the membrane, the solution to Eq.~(\ref{28}) that remains finite for $r\to\infty$ reads 
$\tv_n(r,r';k)\propto\mathrm{K}_n\left(|k|r\right)$, 
where $K_n(x)$ is the modified Bessel function of the second kind~\cite{math}. Inside the nanopore $r<d$ where $\kappa(r)$ is non-uniform, Eq.~(\ref{28}) will be solved within the WKB approximation. First, we note that in the weak-coupling Debye-H\"{u}ckel (DH) approximation where the pore screening parameter equals the bulk value, $\kappa(r)=\kappa_{\rm b}$, the homogeneous solutions are know to be the modified Bessel functions.  Inspired by this point, we will look for solutions of Eq.~(\ref{28}) in the form
\be
\label{29}
\tv_n(r,r';k)=C_1A_n(r)\mathrm{I}_n\left[B_n(r)\right]+C_2A_n(r)\mathrm{K}_n\left[B_n(r)\right],
\ee
where $C_{1,2}$ are integration constants. Due to the linear independence of the Bessel functions ${\rm I}_n(x)$ 
and ${\rm K}_n(x)$, the first and second terms of Eq.~(\ref{29}) should satisfy Eq.~(\ref{28}) independently. Thus, in order to determine the functions $A_n(r)$ and $B_n(r)$, we inject into Eq.~(\ref{28}) only the first term of the ansatz~(\ref{29}). This yields
\bea\label{30}
&&A_n(r)B_n'^2(r)\mathrm{I}''_n\left[B(r)\right]\\
&&+\left\{2A_n'(r)B_n'(r)+A_n(r)B_n''(r)+\frac{A_n(r)B_n'(r)}{r}\right\}\nonumber\\
&&\hspace{5mm}\times\mathrm{I}'_n\left[B_n(r)\right]\nonumber\\
&&+\left\{A_n''(r)+\frac{A_n'(r)}{r}-A_n(r)\left[\frac{n^2}{r^2}+p^2(r)\right]\right\}\nonumber\\
&&\hspace{5mm}\times\mathrm{I}_n\left[B_n(r)\right]=0,\nonumber
\eea
where we defined the local screening parameter
\be
\label{31}
p(r)=\sqrt{\kappa^2(r)+k^2}.
\ee
Now, in Eq.~(\ref{30}), we make use of the following equality satisfied by Bessel functions,
\be\label{32}
\mathrm{I}_n''(x)=-\frac{1}{x}\mathrm{I}_n(x)+\left(\frac{n^2}{x^2}+1\right)\mathrm{I}_n(x),
\ee
which finally yields
\bea\label{33}
&&\left\{A_n''(r)+\frac{A_n'(r)}{r}+A_n(r)B_n'^2(r)\left[\frac{n^2}{B_n^2(r)}+1\right]\right.\nonumber\\
&&\left.\hspace{3mm}-A_n(r)\left[\frac{n^2}{r^2}+p^2(r)\right]\right\}\mathrm{I}_n\left[B_n(r)\right]\nonumber\\
&&+\left\{A_n(r)B_n''(r)+2A_n'(r)B_n'(r)+\frac{A_n(r)B_n'(r)}{r}\right.\nonumber\\
&&\left.\hspace{6mm}-\frac{A_n(r)B_n'^2(r)}{B_n(r)}\right\}\mathrm{I}'_n\left[B_n(r)\right]=0.
\eea

At this stage, we note that the ansatz of Eq.~(\ref{29}) contains two functions that cannot be determined uniquely by the single Eq.~(\ref{28}) or~(\ref{33}). Thus, we have to impose an additional relation between the functions $A_n(r)$ and $B_n(r)$. Inspired by a strategy previously used in the WKB solution of the Schr\"{o}dinger equation in cylindrical coordinates~\cite{WKB1}, we set the bracket term of Eq.~(\ref{33}) proportional to $\mathrm{I}'_n\left[B_n(r)\right]$ to zero,
\be
\label{34}
\frac{B_n''(r)}{B_n'(r)}-\frac{B_n'(r)}{B_n(r)}+\frac{2A_n'(r)}{A_n(r)}+\frac{1}{r}=0.
\ee
The integration of Eq.~(\ref{34}) yields the amplitude of the Green's function Eq.~(\ref{29}) in the form
\be
\label{35}
A_n(r)=\sqrt{\frac{B_n(r)}{rB_n'(r)}}.
\ee
The second bracket term of Eq.~(\ref{33}) being zero, we are left with the equality
\bea\label{36}
&&A_n''(r)+\frac{A_n'(r)}{r}\\
&&+\left\{B_n'^2(r)\left[\frac{n^2}{B_n^2(r)}+1\right]-\left[\frac{n^2}{r^2}+p^2(r)\right]\right\}A_n(r)=0.\nonumber
\eea
At this point, we introduce the WKB approximation. It consists of assuming that the amplitude 
$A_n(r)$ of the solution in Eq.~(\ref{29}) varies slowly. Thus, we neglect the derivative terms in Eq.~(\ref{36}). This yields 
\be\label{36II}
\frac{\mathrm{d}B_n(r)}{dr}\sqrt{\frac{m^2}{B^2_n(r)}+1}=\sqrt{\frac{n^2}{r^2}+p^2(r)}.
\ee
A direct integration of Eq.~(\ref{36II}) gives
\be\label{37}
\int_{0}^{B_n(r)}\mathrm{d}B_n\sqrt{\frac{n^2}{B_n^2}+1}=\int_0^r\mathrm{d}r'\sqrt{\frac{n^2}{r'^2}+p^2(r')}.
\ee
For $n=0$,  Eq.~(\ref{37}) has the trivial solution
\be\label{37II}
B_0(r)=\int_0^r\mathrm{d}r'p(r').
\ee
In the present model where we will restrict the polymer position to the pore axis ($r_{\rm p}=0$), the component with the ground state mode $n=0$ solely contributes to the self-energy in Eq.~(\ref{13}). Thus, Eq.~(\ref{29}) together with Eqs.~(\ref{35}) and~(\ref{37II}) complete the calculation of the homogeneous solutions to Eq.~(\ref{15}). However, in order to show that the modes $n\neq0$ vanish in the mid-pore limit $r_{\rm p}\to0$, we need to complete the present calculation for finite $n$. 

For $n\neq0$, the integrals on both sides of Eq.~(\ref{37}) diverge at their lower bound. By regularizing Eq.~(\ref{37}), this ultraviolet (UV) divergence can be avoided. To this end, we first integrate Eq.~(\ref{36II}) between $r_i$ and $r$ to get
\be\label{37IV}
\int_{B_n(r_i)}^{B_n(r)}\mathrm{d}B_n\sqrt{\frac{n^2}{B_n^2}+1}=\int_{r_i}^r\mathrm{d}r'\sqrt{\frac{n^2}{r'^2}+p^2(r')}.
\ee
Next, based on Eq.~(\ref{37IV}), we note that
\be\label{37III}
B_n(r)\approx p(r)r, \hspace{1cm}\mbox{for}\hspace{2mm}r\to0.
\ee
Evaluating the integral on the l.h.s. of Eq.~(\ref{37IV}), taking the limit $r_i\to0$, and using Eq.~(\ref{37III}), one finally gets
\bea\label{37V}
f\left[B_n(r)/n\right]&=&\lim_{r_i\to0}\left\{\frac{1}{n}\int_{r_i}^r\mathrm{d}r'\sqrt{\frac{n^2}{r'^2}+p^2(r')}\right.\nonumber\\
&&\hspace{1cm}\left.+f\left[p(r_i)r_i/n\right]\right\},
\eea
where we defined the auxiliary function
\be
\label{37VI}
f(x)=\sqrt{1+x^2}-\ln\left(x^{-1}+\sqrt{1+x^{-2}}\right).
\ee
Equation (\ref{37V}) is identical to Eq.~(\ref{37}); we simply subtracted the same ultraviolet divergent quantity from both sides of the equality. In the limit $n\to0$, Eq.~(\ref{37V}) naturally yields Eq.~(\ref{37II}). For Fourier components with finite $n$, the calculation of the function $B_n(r)$ from Eq.~(\ref{37V}) necessitates the numerical inversion of the function $f(x)$.

\subsubsection{Particular solution of the kernel Eq.~(\ref{15})}
\label{part}

Based on the previously derived homogeneous solutions to Eq.~(\ref{15}), we calculate here the particular solution of this equation for ions located in the pore, i.e. $r'<d$. To this end, we impose first the finiteness of the Green's function~(\ref{29}) at $r=0$ and $r\to\infty$. Then, we take into account the absence of ions in the membrane, i.e. $\kappa(r>d)=0$. Consequently, the general solution to Eq.~(\ref{15}) can be expressed as
\bea\label{39}
\tv_n(r,r';k)&=&c_1A_n(r)\mathrm{I}_n\left[B_n(r)\right]\theta(r'-r)\nonumber\\
&&+A_n(r)\left\{c_2\mathrm{I}_n\left[B_n(r)\right]+c_3\mathrm{K}_n\left[B_n(r)\right]\right\}\nonumber\\
&&\hspace{1.1cm}\times\theta(r-r')\theta(d-r)\nonumber\\
&&+c_4\mathrm{K}_n\left(|k|r\right)\theta(r-d).
\eea
In order to determine the integration constants $c_i$, we impose now the boundary conditions of Eqs.~(\ref{19}) - (\ref{22}) to Eq.~(\ref{39}). After long but straightforward algebra, the Green's function finally takes the form
\bea\label{40}
\tv_n(r,r';k)&=&4\pi\ell_{\rm B}A_n(r_<)A_n(r_>)\mathrm{I}_n\left[B_n(r_<)\right]\\
&&\times\left\{\mathrm{K}_n\left[B_n(r_>)\right]+\frac{G_n(k)}{T_n(k)}\mathrm{I}_n\left[B_n(r_>)\right]\right\}.\nonumber
\eea
In Eq.~(\ref{40}), we used the radial variables
\be\label{41}
r_<=\mathrm{min}(r,r')\;;\hspace{5mm}r_>=\mathrm{max}(r,r'), 
\ee
and introduced the auxiliary functions taking into account the nanopore geometry,
\bea
\label{42}
G_n(k)&=&A_n'(d)\mathrm{K}_n\left(|k|d\right)\mathrm{K}_n\left[B_n(d)\right]\\
&&+A_n(d)\mathrm{K}_n\left(|k|d\right)B_n'(d)\mathrm{K}'_n\left[B_n(d)\right]\nonumber\\
&&-\gamma |k|A_n(d)\mathrm{K}'_n\left(|k|d\right)\mathrm{K}_n\left[B_n(d)\right];\nonumber\\
\label{43}
T_n(k)&=&-A_n'(d)\mathrm{K}_n\left(|k|d\right)\mathrm{I}_n\left[B_n(d)\right]\\
&&-A_n(d)\mathrm{K}_n\left(|k|d\right)B_n'(d)\mathrm{I}'_n\left[B_n(d)\right]\nonumber\\
&&+\gamma |k|A_n(d)\mathrm{K}'_n\left(|k|d\right)\mathrm{I}_n\left[B_n(d)\right],\nonumber
\eea
with the dielectric contrast parameter $\gamma=\e_{\rm m}/\e_{\rm w}$.

\subsubsection{Computing the polymer self-energy $\Delta\Omega_{\rm s}(r_{\rm p}=0;l_{\rm p})$}
\label{polsel}

Using the Fourier-transformed Green's function of Eq.~(\ref{40}), we evaluate now the mid-pore value of the self-energy in Eq.~(\ref{13}), {\it i.e.} $\Delta\Omega_{\rm s}(r_{\rm p}\to0,l_{\rm p})$. According to Eq.~(\ref{13}), this requires the evaluation of the following limit,
\bea\label{44}
&&\lim_{r_{\rm p}\to0}\left[\tv_n(r_{\rm p},r_{\rm p};k)-\tv_{{\rm b},n}(r_{\rm p},r_{\rm p};k)\right]=\nonumber\\
&&4\pi\ell_{\rm B}\lim_{r_{\rm p}\to0}\left\{A_n^2(r_{\rm p})\mathrm{I}_n\left[B_n(r_{\rm p})\right]\mathrm{K}_n\left[B_n(r_{\rm p})\right]\right.\nonumber\\
&&\hspace{1.7cm}\left.-\mathrm{I}_n(p_{\rm b}r_{\rm p})\mathrm{K}_n(p_{\rm b}r_{\rm p})\right\}\nonumber\\
&&+4\pi\ell_{\rm B}\frac{G_n(k)}{T_n(k)}\lim_{r_{\rm p}\to0}A_n^2(r_{\rm p})\mathrm{I}_n^2\left[B_n(r_{\rm p})\right],
\eea
where we used the bulk limit of the Green's function of Eq.~(\ref{40})
\be
\label{45}
\tv_{{\rm b},n}(r,r';k)=4\pi\ell_{\rm B}K_n(p_{\rm b}r_>)I_n(p_{\rm b}r_<),
\ee
with $p_{\rm b}=\sqrt{\kappa_{\rm b}^2+k^2}$. We now note that according to Eqs.~(\ref{35}) and~(\ref{37III}), one has 
$A_n(r_{\rm p}\to0)=1$ and $\mathrm{I}_n\left[B_n(r_{\rm p}\to0)\right]=\delta_{n0}$, where $\delta_{n0}$ 
stands for the Kronecker delta function. Using these equalities, the first limit on the r.h.s. of Eq.~(\ref{44}) becomes 
$-4\pi\ell_{\rm B}\ln\left[p(0)/p_{\rm b})\right]\delta_{n0}$. This shows that in the mid-pore limit, the 
ground state mode $n=0$ solely brings a finite contribution to the polymer self-energy. Finally, using Eqs.~(\ref{35}) and~(\ref{37II}) in order to simplify Eqs.~(\ref{42}) and~(\ref{43}), the mid-pore value of the self-energy~(\ref{13}) takes the form
\bea\label{46}
\beta\Delta\Omega_{\rm s}(0,l_{\rm p})&=&l_{\rm p}\ell_{\rm B}\tau^2\int_{-\infty}^\infty\mathrm{d}k\frac{2\sin^2(kl_{\rm p}/2)}{\pi l_{\rm p}k^2}\\
&&\hspace{1.9cm}\times\left\{-\ln\left[\frac{p(0)}{p_{\rm b}}\right]+\frac{Q(k)}{P(k)}\right\},\nonumber
\eea
where we introduced the auxiliary functions
\bea\label{47}
Q(k)&=&2p^3(d)dB_0(d)\mathrm{K}_0\left(|k|d\right)\mathrm{K}_1\left[B_0(d)\right]\\
&&-2\gamma |k|dp^2(d)B_0(d)\mathrm{K}_1\left(|k|d\right)\mathrm{K}_0\left[B_0(d)\right]\nonumber\\
&&-\left[p^3(d)d-p^2(d)B_0(d)-\kappa(d)\kappa'(d)dB_0(d)\right]\nonumber\\
&&\hspace{3mm}\times\mathrm{K}_0\left(|k|d\right)\mathrm{K}_0\left[B_0(d)\right];\nonumber\\
\label{48}
P(k)&=&2p^3(d)dB_0(d)\mathrm{K}_0\left(|k|d\right)\mathrm{I}_1\left[B_0(d)\right]\\
&&+2\gamma |k|dp^2(d)B_0(d)\mathrm{K}_1\left(|k|d\right)\mathrm{I}_0\left[B_0(d)\right]\nonumber\\
&&+\left[p^3(d)d-p^2(d)B_0(d)-\kappa(d)\kappa'(d)dB_0(d)\right]\nonumber\\
&&\hspace{3mm}\times\mathrm{K}_0\left(|k|d\right)\mathrm{I}_0\left[B_0(d)\right].\nonumber
\eea
The MF component of Eq.~(\ref{27}) and the self-energy in Eq.~(\ref{46}) complete 
the calculation of the polymer grand potential of Eq.~(\ref{1}). The beyond-MF 
polymer-pore interactions embodied in these equations are throughly investigated in Section~\ref{res}.

\section{Results}
\label{res}

Here, we investigate charge correlation effects on the polymer-pore interactions. In the following, we will first focus on the thermodynamic limit $l_{\rm p}\to\infty$ corresponding to the case where the polymer portion in the pore is long enough, i.e. $\kappa_{\rm b}l_{\rm p}\gg1$. In this limit, the sinusoidal function in Eq.~(\ref{46}) becomes a Dirac delta function, and the polymer self-energy simplifies to
\begin{widetext}
\be\label{49}
\beta\Delta\Omega_{\rm s}(0;l_{\rm p})=-l_{\rm p}\ell_{\rm B}\tau^2\ln\left[\frac{\kappa(0)}{\kappa_{\rm b}}\right]+
l_{\rm p}\ell_{\rm B}\tau^2\frac{2\kappa^2(d)dB(d)\mathrm{K}_1\left[B(d)\right]-\left\{\kappa^2(d)d-\left[\kappa(d)+\kappa'(d)d\right]B(d)\right\}\mathrm{K}_0\left[B(d)\right]}{2\kappa^2(d)dB(d)\mathrm{I}_1\left[B(d)\right]+\left\{\kappa^2(d)d-\left[\kappa(d)+\kappa'(d)d\right]B(d)\right\}\mathrm{I}_0\left[B(d)\right]},
\ee
\end{widetext}
where we introduced the infrared (IR) limit of Eq.~(\ref{37II}), 
\be\label{50}
B(r)=\lim_{k\to0}B_0(r)=\int_0^r\mathrm{d}r'\kappa(r').
\ee
In Eq.~(\ref{49}), the negative term is logarithmically proportional to the ratio of the salt densities in the pore and the reservoir. This component accounts for the ionic excess induced by the cation attraction into the negatively charged pore. The resulting salt screening excess lowers the polymer free energy with respect to the bulk reservoir and favours the polymer capture by the pore. The second positive term arising from polymer-image charge interactions prevents the polymer from penetrating the pore. The competition between these two components and the repulsive MF potential of Eq.~(\ref{27}) will be throughly scrutinized for monovalent and multivalent solutions in Sections~\ref{mon} and~\ref{mul}, respectively. In Section~\ref{finlp}, we will also compute the polymer self-energy of Eq.~(\ref{46}) at finite polymer penetration length $l_{\rm p}$ in order to evaluate the grand potential landscape of the polymer during its capture by the nanopore.

\subsection{Symmetric monovalent electrolytes}
\label{mon}

\begin{figure}
\includegraphics[width=1.0\linewidth]{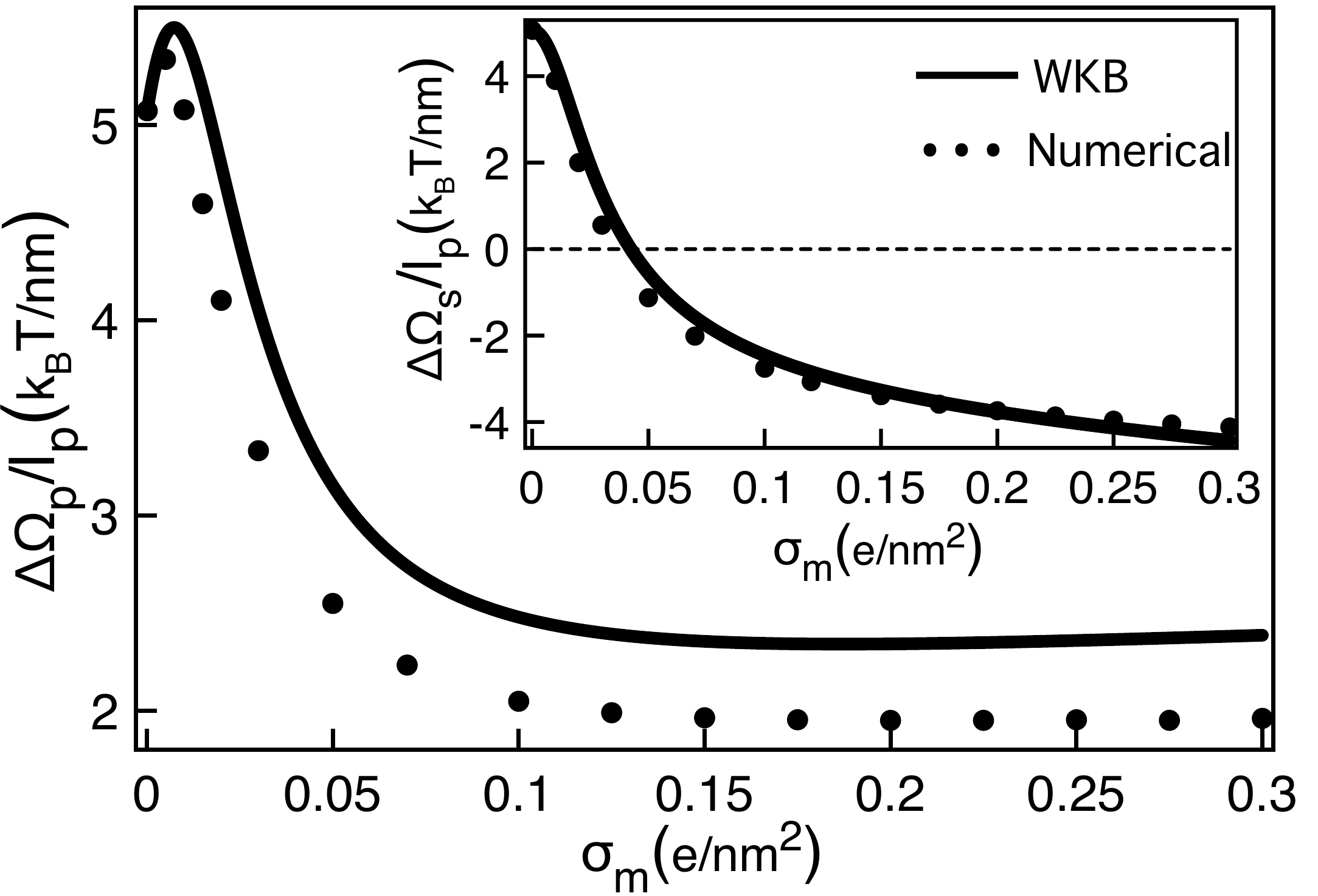}
\caption{(Color online) Thermodynamic limit $l_{\rm p}\to\infty$ of the total grand potential 
$\Delta\Omega_{\rm p}$ (main plot) and the polymer self-energy $\Delta\Omega_{\rm s}$ (inset) versus membrane charge 
$\sigma_{\rm m}$ in a monovalent solution of bulk density $\rho_{\rm b}=0.01$ M. The nanopore radius is $d=3$ nm. 
Solid curves are obtained from Eqs.~(\ref{27}) and~(\ref{49}) and the dots from the numerical solution of Eqs.~(\ref{12}) - (\ref{15}).}
\label{fig2}
\end{figure}

We consider here a symmetric monovalent electrolyte of type NaCl with the ions of valency $q_+=-q_-=1$ and bulk densities 
$\rho_{{\rm b}+}=\rho_{{\rm b}-}=\rho_{\rm b}$. In Fig.~\ref{fig2}, the curves illustrate the effect of the membrane charge on the polymer 
self-energy of Eq.~(\ref{49}) (inset) and the total grand potential of Eq.~(\ref{1}) obtained with the inclusion of the MF 
component of Eq.~(\ref{27}) (main plot). The dots display the exact result obtained from Eqs.~(\ref{12}) - (\ref{13}) with the numerical solution of Eqs.~(\ref{14}) - (\ref{15}) (see Appendix~\ref{apx1}). One notes the reasonably good agreement between the numerical solution and the WKB approach. The WKB result overestimates the total grand potential by $\sim0.5$ $k_{\rm B}T/\mbox{nm}$ but it can accurately capture the effect of the membrane charge.

To gain an analytical insight into the behaviour of the curves in Fig~\ref{fig2}, we switch to the pure Donnan approximation and set $\phi_{\rm m}(r)=\phi_{\rm D}$ and $\kappa(r)=\kappa_{\rm D}$. The grand potential components Eqs.~(\ref{27}) and~(\ref{49}) become
\bea\label{51}
\beta\Omega_{\rm MF}&\approx&-l_{\rm p}\tau\phi_{\rm D};\\
\label{52}
\beta\Delta\Omega_{\rm s}&\approx& l_{\rm p}\ell_{\rm B}\tau^2\left[-\ln\left(\frac{\kappa_{\rm D}}{\kappa_{\rm b}}\right)+\frac{\mathrm{K}_1(\kappa_{\rm D}d)}{\mathrm{I}_1(\kappa_{\rm D}d)}\right].
\eea
For symmetric electrolytes, the Donnan potential follows from the solution of Eq.~(\ref{23}) as 
$\phi_{\rm D}=-\ln\left(t+\sqrt{t^2+1}\right)$, with the auxiliary parameter $t=4/(\kappa_{\rm b}^2\mu d)$, 
where $\mu=1/(2\pi\ell_{\rm B}\sigma_{\rm m})$ is the Gouy-Chapman (GC) length. 
From Eq.~(\ref{25}), the screening parameter follows as $\kappa_{\rm D}=(1+t^2)^{1/4}\kappa_{\rm b}$. We first focus on the DH regime of weakly charged membranes, i.e. $\kappa_{\rm b}\mu\gg1$. Using the equations above, we Taylor expand the grand potential components of Eqs.~(\ref{51}) and~(\ref{52}) in terms of the membrane charge $\sigma_{\rm m}$. To the leading order in $\sigma_{\rm m}$, this yields
\bea\label{53}
\beta\Omega_{\rm MF}&\approx&\frac{l_{\rm p}\tau\sigma_{\rm m}}{d\rho_{\rm b}};\\
\label{54}
\beta\Delta\Omega_{\rm s}&\approx&l_{\rm p}\ell_{\rm B}\tau^2\left[\frac{\mathrm{K}_1(\kappa_{\rm b}d)}{\mathrm{I}_1(\kappa_{\rm b}d)}-\frac{1+\mathrm{I}_1^2(\kappa_{\rm b}d)}{\mathrm{I}_1^2(\kappa_{\rm b}d)}\frac{\sigma_{\rm m}^2}{4d^2\rho_{\rm b}^2}\right].\nonumber\\
\eea
In agreement with the inset of Fig~\ref{fig2}, in neutral membranes with $\sigma_{\rm m}=0$, 
where the image charge barrier in Eq.~(\ref{54}) survives only, the self-energy is positive. With the rise of the membrane charge, the negative term resulting from the cation excess takes over the image-charge component and the self-energy becomes attractive. 

We focus now on the total grand potential corresponding to the sum of Eqs.~(\ref{53}) and~(\ref{54}). As the repulsive MF component scales linearly with $\sigma_{\rm m}$, the grand potential initially rises with the membrane charge 
($\sigma_{\rm m}\uparrow\Delta\Omega_{\rm p}\uparrow$). Beyond a characteristic charge $\sigma_{\rm m}^*$, the attractive part of the self-energy quadratic in $\sigma_{\rm m}$ dominates the MF component and lowers the total grand potential 
($\sigma_{\rm m}\uparrow\Delta\Omega_{\rm p}\downarrow$). This non-monotonic 
behaviour is illustrated in the main plot of Fig.~\ref{fig2}. The location of the peak follows from the equality 
$\partial\left(\beta\Omega_{\rm MF}+\beta\Delta\Omega_{\rm s}\right)/\partial\sigma_{\rm m}=0$ as
\be
\label{55}
\sigma^*_{\rm m}=\frac{2d\rho_{\rm b}}{\ell_{\rm B}\tau}\frac{\mathrm{I}_1^2(\kappa_{\rm b}d)}{1+\mathrm{I}_1^2(\kappa_{\rm b}d)}\approx\frac{2d\rho_{\rm b}}{\ell_{\rm B}\tau}, \hspace{5mm}\mbox{for}\hspace{3mm}\kappa_{\rm b}d\gg1.
\ee
This threshold charge diminishes with the polymer charge density $\tau\uparrow\sigma^*_{\rm m}\downarrow$, and rises with the salt concentration $\rho_{\rm b}\uparrow\sigma^*_{\rm m}\uparrow$ and the nanopore radius $d\uparrow\sigma^*_{\rm m}\uparrow$.

Figure \ref{fig2} shows that in the high membrane charge regime $\sigma_{\rm m}\gtrsim0.1$ $\mbox{e/nm}^2$, the total polymer grand potential is weakly affected by the membrane charge. In order to understand this point, we consider the 
GC regime of strong charges $\kappa_{\rm b}\mu\ll1$ and expand Eqs.~(\ref{51}) and~(\ref{52}) in terms of the inverse membrane charge. This reveals the logarithmic behaviour of the grand potential components,
\bea\label{55}
\beta\Omega_{\rm MF}&\approx&l_{\rm p}\tau\ln\left(\frac{2\sigma_{\rm m}}{d\rho_{\rm b}}\right);\\
\label{56}
\beta\Delta\Omega_{\rm s}&\approx&-\frac{l_{\rm p}\ell_{\rm B}\tau^2}{2}\ln\left(\frac{\sigma_{\rm m}}{d\rho_{\rm b}}\right).
\eea
In the case of ds-DNA with charge density $\tau\approx1.75/\ell_{\rm B}$, the slope of the grand potential 
components in Eqs.~(\ref{55}) and~(\ref{56}) practically cancel each other out. This explains the saturation of the 
grand potential in Fig.~\ref{fig2}. 

For the parameters of Fig.~\ref{fig2}, we found that the grand potential is positive and the nanopore repels the ds-DNA at any membrane charge. At this point, the question arises whether the like-charge DNA-pore attraction can ever occur in monovalent solutions. This requires the self-energy of Eq.~(\ref{56}) to dominate the MF component of Eq.~(\ref{55}). Thus, the membrane charge should satisfy the inequality
\be\label{57}
\sigma_{\rm m}>2^{2/(\ell_{\rm B}\tau-2)}\rho_{\rm b}d.
\ee
Deriving the condition above, we assumed that the self-energy of Eq.~(\ref{56}) is negative, {\it i.e.} $\sigma_{\rm m}/(d\rho_{\rm b})>1$. Thus, the validity of Eq.~(\ref{57}) requires the polymer charge density to satisfy $\tau>\tau_{\rm c}=2/\ell_{\rm B}$. 
Since the ds-DNA charge density $\tau\simeq1.75/\ell_{\rm B}$ is below $\tau_{\rm c}$, like-charge DNA-pore attraction cannot occur in monovalent electrolytes. Next, we consider the case of solutions including polyvalent cations.

\subsection{Electrolyte mixtures with polyvalent cations}
\label{mul}

\subsubsection{Polyvalent cation-induced DNA-pore attraction}

\begin{figure}
\includegraphics[width=1.1\linewidth]{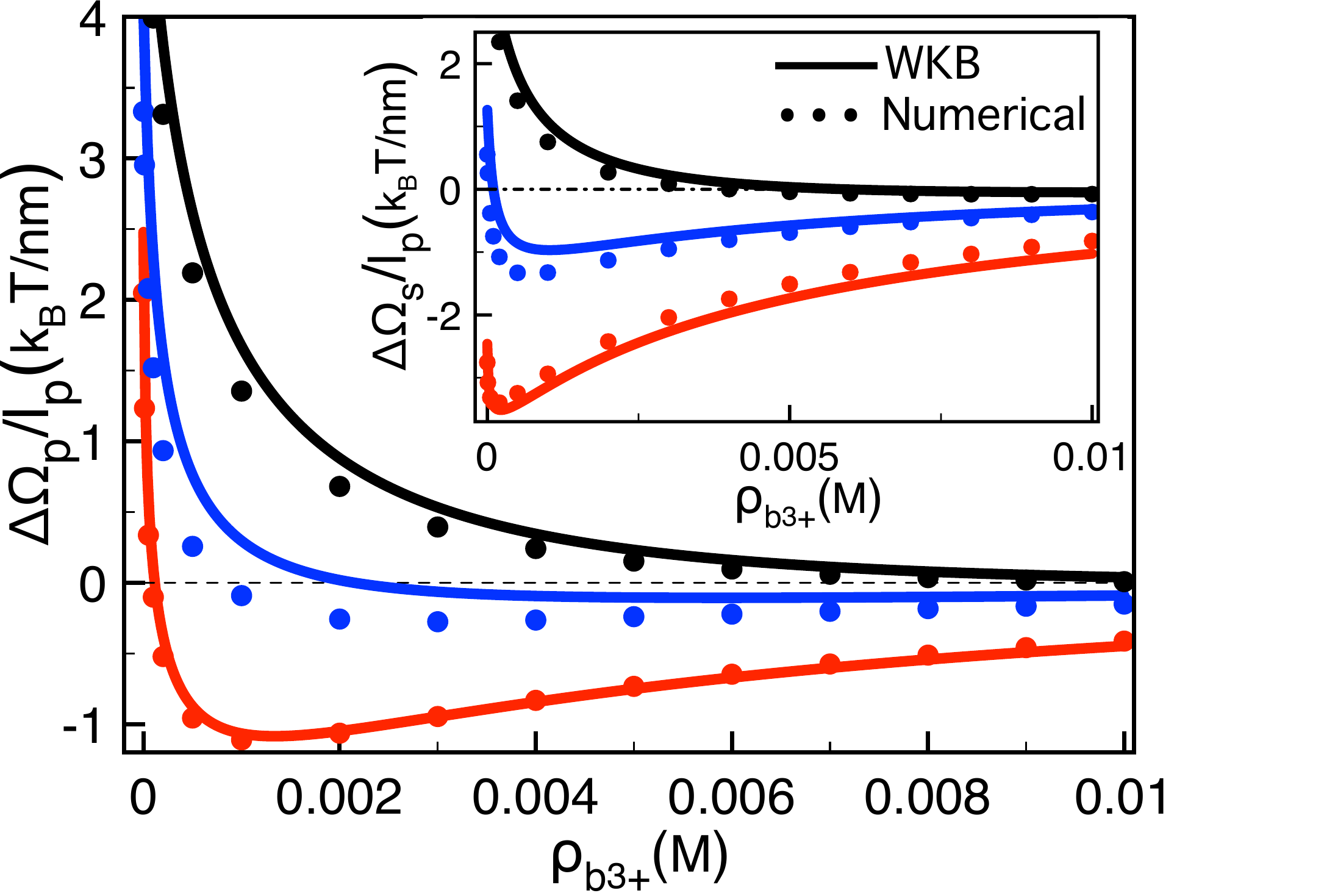}
\caption{(Color online) Thermodynamic limit $l_{\rm p}\to\infty$ of the total grand potential $\Delta\Omega_{\rm p}$ (main plot) and the polymer self-energy $\Delta\Omega_{\rm s}$ (inset) versus trivalent cation density 
$\rho_{b3+}$ in the electrolyte mixture $\mbox{NaCl}+\mbox{SpdCl}_3$ with monovalent cation density $\rho_{{\rm b}+}=0.01$ M. The membrane charge is $\sigma_{\rm m}=0.01$ $\mbox{e/nm}^2$ (black), $\sigma_m=0.03$ $\mbox{e/nm}^2$ (blue), and $\sigma_m=0.1$ $\mbox{e/nm}^2$ (red). The remaining parameters are the same as in Fig.~\ref{fig2}.}
\label{fig3}
\end{figure}

We investigate now polymer-pore interactions in mixed solutions $\mbox{NaCl}+\mbox{XCl}_m$ including the polyvalent cation species $X^{m+}$. First, we consider the electrolyte mixture $\mbox{NaCl}+\mbox{SpdCl}_3$. Figure \ref{fig3} displays the total polymer grand potential (main plot) and the self-energy (inset) against the bulk spermidine ($\mbox{Spd}^{3+}$) concentration at various membrane charges. The comparison of the curves and dots shows that the WKB approach can reproduce the polymer grand potential with reasonably good accuracy.

In order to interpret the grand potential curves, we switch to the Donnan approximation and focus on the DH regime 
$\kappa_{\rm b}\mu\gg1$ of weak membrane charges. On the linear order in the membrane charge density $\sigma_{\rm m}$, the Donnan potential and screening parameter follow from Eqs.~(\ref{23}) and~(\ref{25}) as $\phi_{\rm D}\approx-4/(\kappa_{\rm b}^2\mu d)$ and $\kappa_{\rm D}^2\approx\kappa_{\rm b}^2-4\pi\ell_{\rm B}(m^3-m)\rho_{{\rm b}m+}\phi_{\rm D}$. By substituting these expressions into the grand potential components of Eqs.~(\ref{51}) and~(\ref{52}) and Taylor expanding them, to the leading order one finds
\bea\label{58}
\beta\Omega_{\rm MF}&\approx&\frac{2l_{\rm p}\tau\sigma_{\rm m}}{d\left[2\rho_{{\rm b}+}+(m^2+m)\rho_{{\rm b}m+}\right]};\\
\label{59}
\beta\Delta\Omega_{\rm s}&\approx&l_{\rm p}\ell_{\rm B}\tau^2\left\{\frac{\mathrm{K}_1(\kappa_{\rm b}d)}{\mathrm{I}_1(\kappa_{\rm b}d)}\right.\\
&&\hspace{5mm}\left.-\frac{1+\mathrm{I}_1^2(\kappa_{\rm b}d)}{\mathrm{I}_1^2(\kappa_{\rm b}d)}\frac{(m^3-m)\rho_{{\rm b}m+}\sigma_{\rm m}}{d\left[2\rho_{{\rm b}+}+(m^2+m)\rho_{{\rm b}m+}\right]^2}\right\}.\nonumber
\eea
The negative term of Eq.~(\ref{59}) indicates that the addition of multivalent cations to the monovalent solution lowers the polymer self-energy. This feature is displayed in the inset of Fig.~\ref{fig3}. In particular, at the membrane charge $\sigma_{\rm m}=0.03$ $\mbox{e/nm}^2$ (blue curve), multivalent cations solely remove the image-charge barrier and switch the self-energy from repulsive to attractive. The main plot shows that as a result of this effect, beyond a characteristic membrane charge, the addition of polyvalent cations turns the grand potential from positive to negative and triggers the attraction of the DNA molecule by the like-charged nanopore. This is the key prediction of our theory. Then, due to the denominator of the second term in Eq.~(\ref{59}), the same multivalent cations screen the self-energy.  Figure \ref{fig3} shows that beyond a characteristic $\mbox{Spd}^{3+}$ concentration, this attenuates the magnitude of the polymer self-energy and the attractive grand potential. 

\begin{figure}
\includegraphics[width=1.0\linewidth]{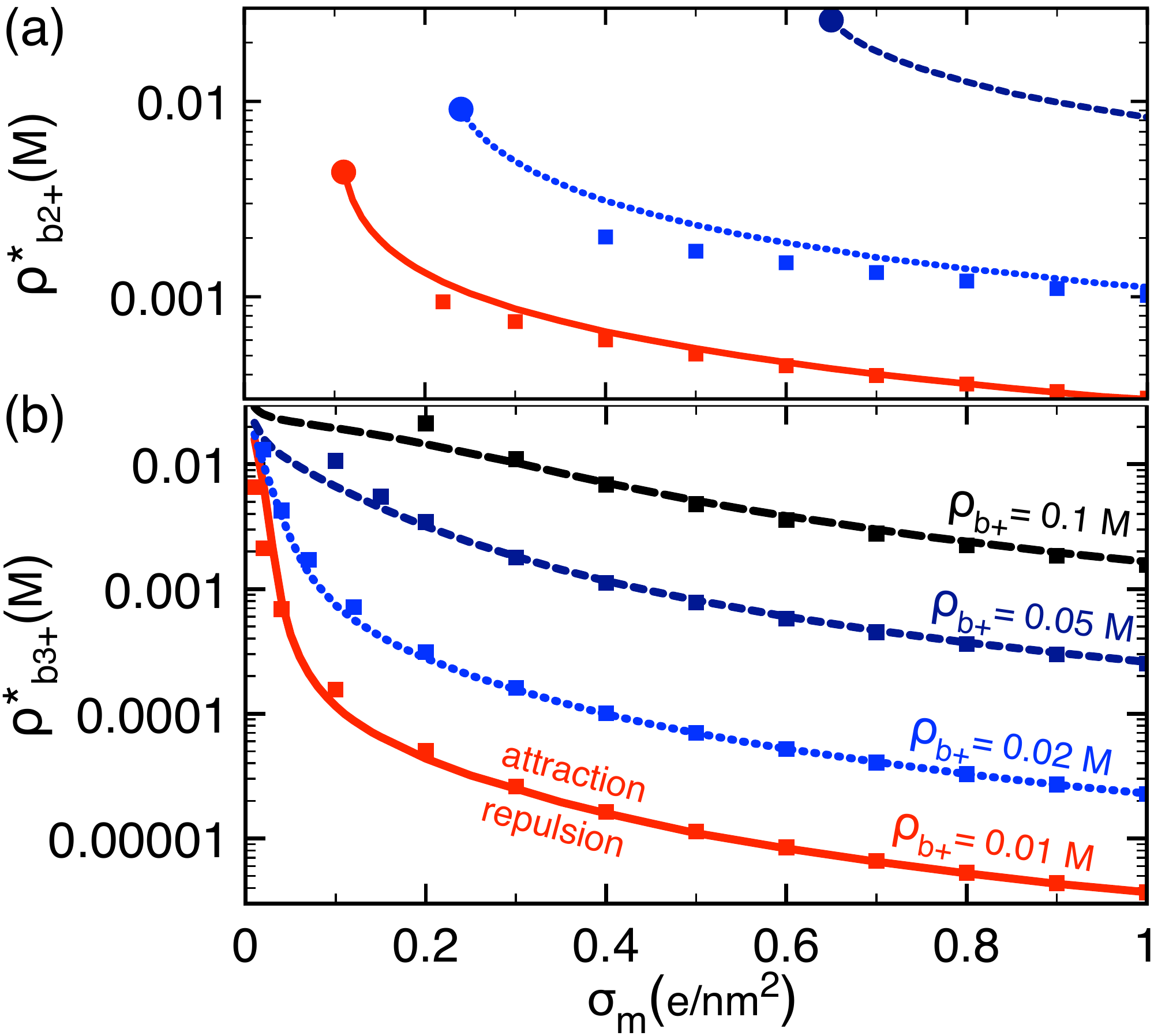}
\caption{(Color online) Phase diagram: critical multivalent cation concentration $\rho^*_{{\rm b}m+}$ versus membrane charge density 
curves splitting the parameter regimes with attractive and repulsive polymer-pore interactions in the electrolyte mixtures (a) $\mbox{NaCl}+\mbox{MgCl}_2$ ($m=2$) and (b) $\mbox{NaCl}+\mbox{SpdCl}_3$ ($m=3$). The monovalent cation concentration 
$\rho_{{\rm b}+}$ is indicated above each curve. The other parameters are the same as in Fig.~\ref{fig2}. 
The square symbols correspond to the scaling law of Eq.~(\ref{63}) with the fitting parameter $c_m=4.0$ in (a) and $c_m=5.2$ in (b).}
\label{fig4}
\end{figure}

\subsubsection{Effect of membrane charge, monovalent salt concentration, and cation valency}

According to Eq.~(\ref{59}), the magnitude of the attractive self-energy component is lowered by the reduction of the membrane charge $\sigma_{\rm m}$ or the cation valency $m$, and the rise of the monovalent salt density $\rho_{{\rm b}+}$. Thus, in order for the net interaction to remain attractive, this has to be compensated by a larger multivalent cation concentration $\rho_{{\rm b}m+}$. This effect is illustrated in Figs.~\ref{fig4} (a) and (b) respectively for $\mbox{Mg}^{2+}$ and $\mbox{Spd}^{3+}$ cations. The diagrams display the critical multivalent cation concentration $\rho^*_{{\rm b}m+}$ where polymer-pore interactions become attractive versus the membrane charge $\sigma_{\rm m}$ at various monovalent salt concentration values $\rho_{{\rm b}+}$. One notes that the critical multivalent cation density increses with decreasing membrane charge ($\sigma_{\rm m}\downarrow\;\rho^*_{{\rm b}m+}\uparrow$) or increasing monovalent salt  density 
($\rho_{{\rm b}+}\uparrow\;\rho^*_{{\rm b}m+}\uparrow$). Moreover, the comparison of Figs.~(\ref{fig4}) (a) and (b) shows that the critical $\mbox{Mg}^{2+}$ density for the occurrence of DNA-pore attraction is more than an order of magnitude higher than the critical $\mbox{Spd}^{3+}$ density. One also notes that in the $\mbox{NaCl}+\mbox{MgCl}_2$ liquid, the critical curves end at a critical point (dots) where the like-charge attraction phase disappears.

We derive now a scaling law that can explain the trend of the critical lines in Fig.~\ref{fig4}. In the GC regime 
$\kappa_{\rm b}\mu\ll1$, the Donnan potential and screening parameter follow from Eqs.~(\ref{23}) and~(\ref{25}) as $\phi_{\rm D}\approx m^{-1}\ln\left[m\rho_{{\rm b}m+}d/(2\sigma_{\rm m})\right]$ and $\kappa_D^2\approx8\pi\ell_{\rm B}m\sigma_{\rm m}/d$. Injecting these equalities into Eqs.~(\ref{51}) and~(\ref{52}) and expanding the result, one finds
\bea
\label{60}
\beta\Omega_{\rm MF}&\approx&\frac{l_{\rm p}\tau}{m}\ln\left[\frac{2\sigma_{\rm m}}{md\rho_{{\rm b}m+}}\right];\\
\label{61}
\beta\Delta\Omega_{\rm s}&\approx&-\frac{\ell_{\rm B}l_{\rm p}\tau^2}{2}\ln\left[\frac{2md^{-1}\sigma_{\rm m}}
{2\rho_{{\rm b}+}+(m^2+m)\rho_{{\rm b}m+}}\right].
\eea
According to Eqs.~(\ref{60}) and~(\ref{61}), the total grand potential becomes attractive in the membrane charge regime corresponding to
\be\label{62}
\sigma_{\rm m} > \frac{d}{2}\left\{\frac{\left[2\rho_{{\rm b}+}+(m^2+m)\rho_{{\rm b}m+}\right]^{m\ell_{\rm B}\tau}}
{m^{m\ell_{\rm B}\tau+2}\rho^2_{{\rm b}m+}}\right\}^{1/(m\ell_{\rm B}\tau-2)}.
\ee
For dilute polyvalent cations, Eq.~(\ref{62}) indicates that polymer-pore attraction occurs in the regime 
$\rho_{{\rm b}m+}>\rho^*_{{\rm b}m+}$ with the critical concentration
\be\label{63}
\rho_{{\rm b}m+}^*\approx c_{m}\;d^{m \ell_{\rm B} \tau/2-1}\rho_{{\rm b}+}^{m \ell_{\rm B} \tau/2}
\sigma_{\rm m}^{-(m \ell_{\rm B}\tau/2-1)}
\ee
and the adimensional parameter $c_{m}=2m^{-m \ell_{\rm B} \tau/2-1}$. We found that Eq.~(\ref{63}) derived within the Donnan approximation underestimates the critical concentration. However, by fitting the parameter $c_{m}$ once for each of the graphs in Fig.~\ref{fig4}, in the corresponding GC regime, Eq.~(\ref{63}) can correctly reproduce the alteration of the critical concentration by the membrane charge and monovalent salt density (square symbols). Again, we emphasize that Eq.~(\ref{63}) is proposed here as a 
\textit{scaling ansatz} that can be useful for translocation experiments rather than an accurate asymptotic law. Interestingly, Eq.~(\ref{63}) predicts the decrease of the critical  multivalent cation density with the pore size, i.e. $d\downarrow\rho^*_{{\rm b}m+}\downarrow$. The corresponding pore confinement effects will be investigated in the next part.

The validity of Eq.~(\ref{62}) requires the GC self-energy~(\ref{61}) to be negative. Together with Eq.~(\ref{62}), this implies that like-charge polymer-pore attraction can occur only in the polymer charge density regime $\tau>\tau_{\rm c}=2/(m\ell_{\rm B})$. 
In solutions including polyvalent cations ({\it i.e.} $m\geq2$), this condition is indeed satisfied by the characteristic charge density of ds-DNA molecules $\tau\approx1.75/\ell_{\rm B}$.

\begin{figure}
\includegraphics[width=1.0\linewidth]{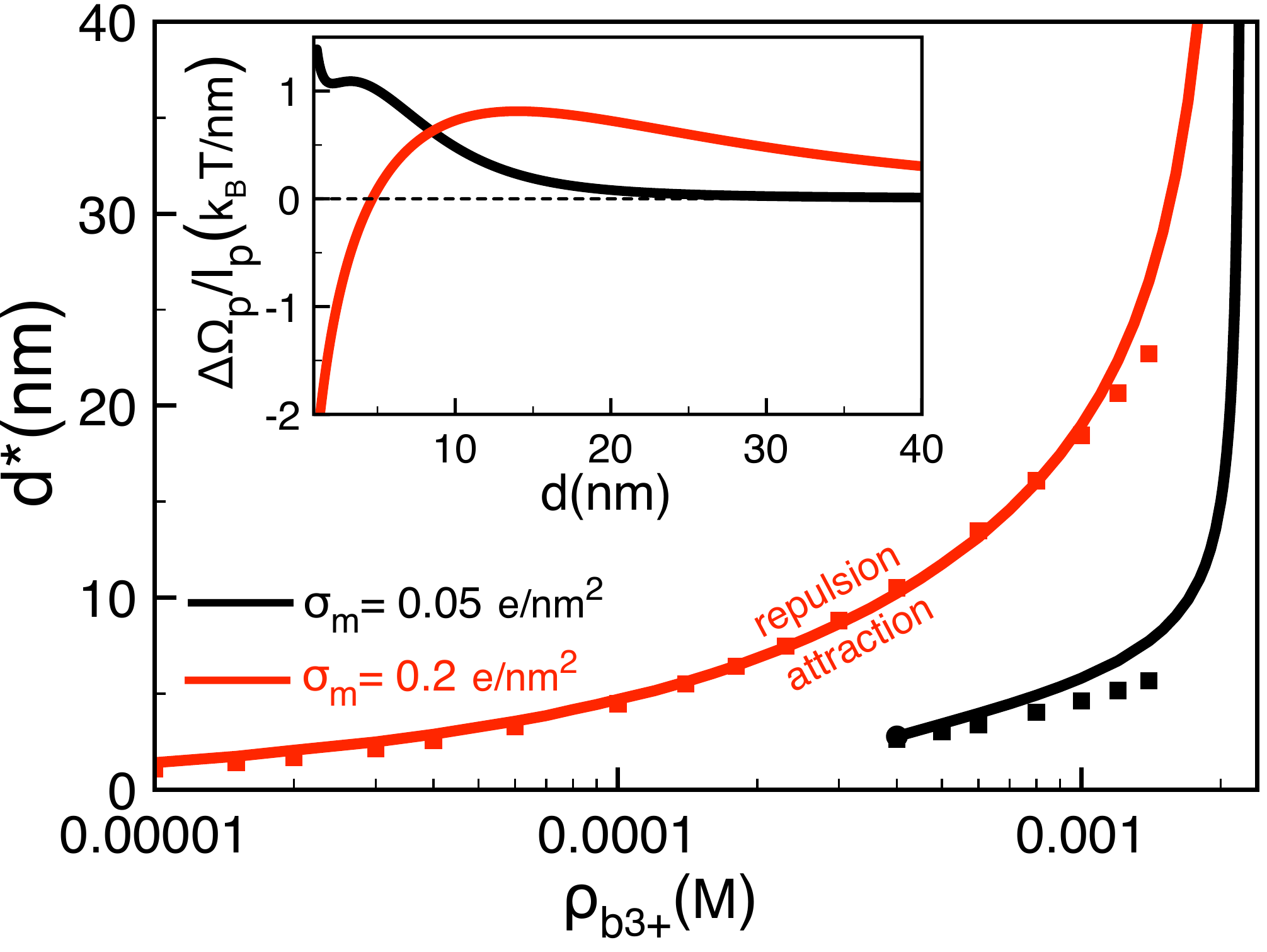}
\caption{(Color online) Main plot: critical pore radius $d^*$ where polymer-pore interactions turn from repulsive to attractive against the $\mbox{Spd}^{3+}$ concentration. Inset: total polymer grand potential versus the pore radius $d$ at the  $\mbox{Spd}^{3+}$ concentration $\rho_{b3+}=10^{-4}$ M. The monovalent cation density is $\rho_{{\rm b}+}=0.01$ M. The membrane charge is
$\sigma_{\rm m}=0.05$ $\mbox{e/nm}^2$ (black) and $\sigma_{\rm m}=0.2$ $\mbox{e/nm}^2$ (red). The remaining parameters are the same as in Fig.~\ref{fig2}. The square symbols are from the scaling law of Eq.~(\ref{64}) with the fitting parameter $c'_{\rm m}=0.6$.}
\label{fig5}
\end{figure}

\subsubsection{Effect of pore confinement on polymer-pore interactions}

In this section we consider the effect of the pore confinement.  Figure \ref{fig5} displays the critical pore radius where the polymer grand potential becomes attractive against the $\mbox{Spd}^{3+}$ density. The location of the attraction phase below the critical lines indicates that despite the presence of the image-charge barrier, confinement favours the attraction of the polymer by the like-charged pore. This point is also illustrated in the inset. In weakly charged pores (black curve), due to the image-charge barrier, the grand potential becomes more repulsive with decreasing pore size ($d\downarrow\Delta\Omega_{\rm p}\uparrow$). In strongly charged pores (red curve), the interaction is repulsive at large pore radii but becomes attractive below a characteristic pore radius ($d\downarrow\Delta\Omega_{\rm p}\downarrow$).

Comparing Eqs.~(\ref{59}) and~(\ref{61}), one notes that the transition from the DH to the GC regime through the increase of the membrane charge removes the image-charge barrier and the self-energy becomes purely 
attractive. In this strong membrane charge regime,  the attractive self-energy of Eq.~(\ref{61}) takes over the repulsive MF component of Eq.~(\ref{60}) if the pore radius is lowered below the critical value
\be
\label{64}
d^*\approx c'_{m} \left(\rho_{{\rm b}m+}\right)^{2/(m\ell_{\rm B}\tau-2)}
\left(\rho_{{\rm b}+}\right)^{-m\ell_{\rm B} \tau/(m \ell_{\rm B} \tau-2)}\sigma_{\rm m}.
\ee
This explains the enhancement of like-charge attraction by pore confinement at strong enough membrane charge. Furthermore, Fig.~\ref{fig5} shows that with a single fitting parameter $c'_m$, the scaling law of Eq.~(\ref{64}) can accurately reproduce the increase of the critical radius with the polyvalent cation density 
$\rho_{{\rm b}m+}\uparrow d^*\uparrow$ and the membrane charge $\sigma_{\rm m}\uparrow d^*\uparrow$. In Fig.~\ref{fig5}, the validity of Eq.~(\ref{64}) at low pore radii can be explained by Eq.~(\ref{23}). This relation shows that the reduction of the pore size and the increment of the membrane charge are equivalent as both effects enhance the electrostatic potential in the pore.


\subsubsection{Polymer grand potential profile during the capture regime}
\label{finlp}

\begin{figure}
\includegraphics[width=1.1\linewidth]{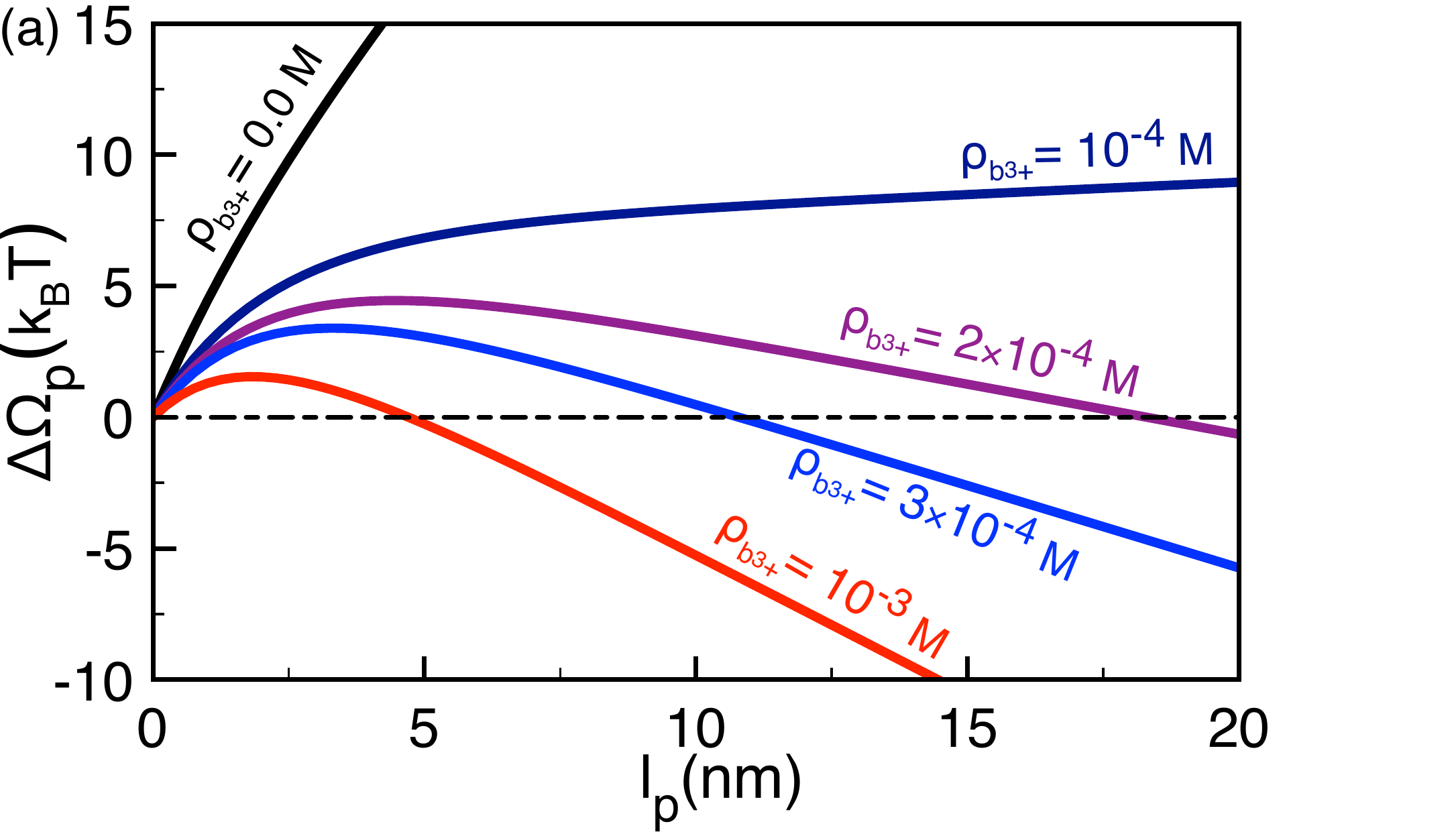}
\includegraphics[width=1.1\linewidth]{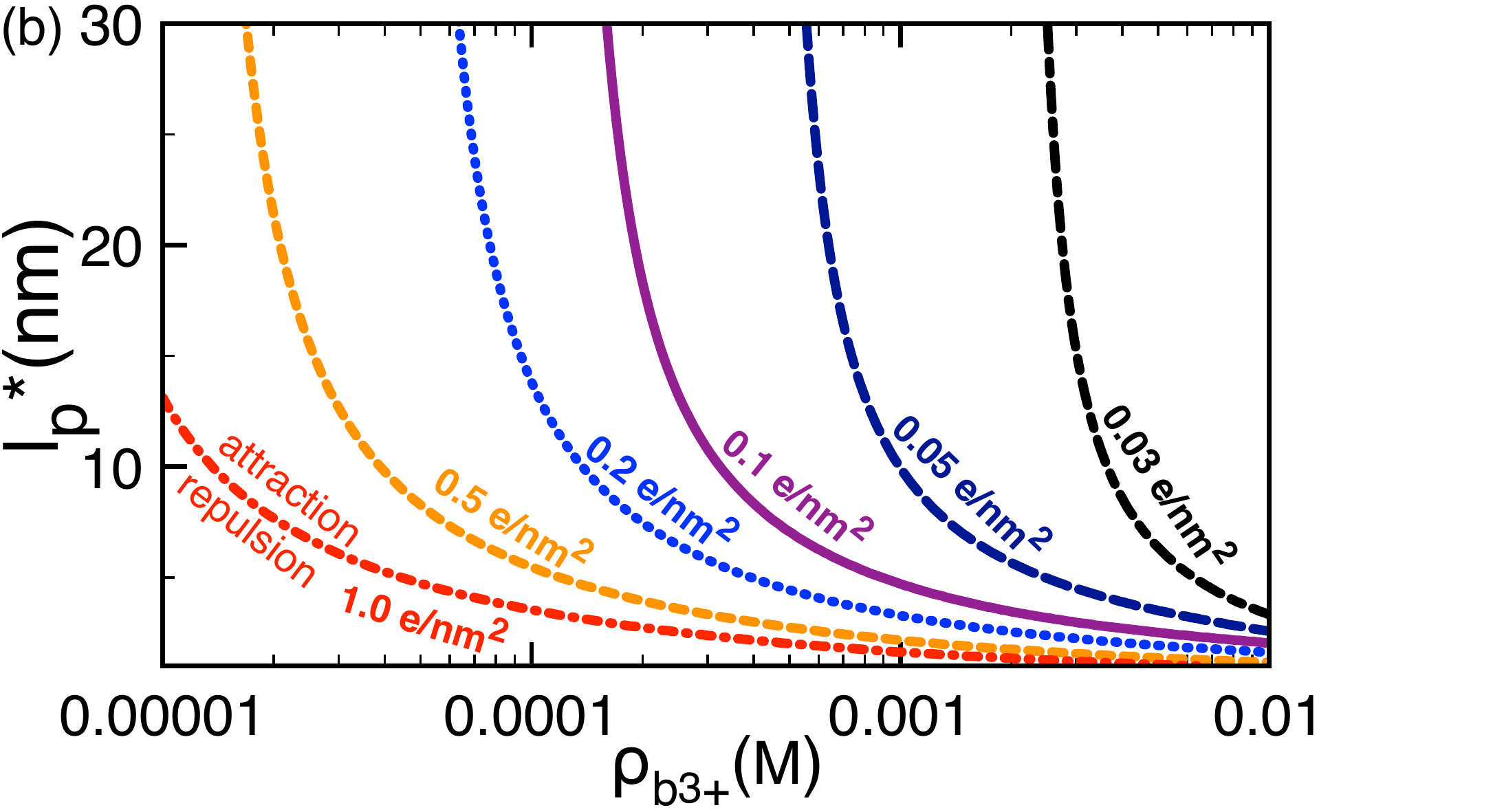}
\caption{(Color online)  (a) The total grand potential $\Delta\Omega_{\rm p}$ from Eqs.~(\ref{27}) and~(\ref{46}) 
versus the length $l_{\rm p}$ of the polymer portion in the pore at various $\mbox{Spd}^{3+}$ densities $\rho_{{\rm b}3+}$. 
The membrane charge density is $\sigma_{\rm m}=0.1$ $\mbox{e/nm}^2$. (b) Critical penetration length $l_{\rm p}^*$ where the grand potential becomes attractive against the $\mbox{Spd}^{3+}$ density at different membrane charge densities $\sigma_{\rm m}$. The other parameters are the same as in Fig.~\ref{fig2}.}
\label{fig6}
\end{figure}

Finally, we investigate the electrostatic barrier experienced by the polymer during its capture into the pore. This necessitates the evaluation of the grand potential $\Delta\Omega_{\rm p}$ at finite polymer length $l_{\rm p}$. At this point, the WKB solution of 
Eq.~(\ref{46}) becomes crucial;  due to the extensive memory requirement, the exact numerical evaluation of the polymer self-energy from Eqs.~(\ref{13}) and~(\ref{ap7}) is simply intractable. Figure \ref{fig6}(a) displays the grand potential profile versus the length 
$l_{\rm p}$ at various $\mbox{Spd}^{3+}$ concentration values. In the monovalent NaCl solution where polymer-pore interactions are driven by the MF component of Eq.~(\ref{27}) proportional to the length $l_{\rm p}$, the repulsive grand potential rises in a quasilinear fashion (black curve).  In the $\mbox{Spd}^{3+}$ density regime $\rho_{{\rm b}3+}>10^{-4}$ M, the grand potential increases
 ($l_{\rm p}\uparrow\Delta\Omega_{\rm p}\uparrow$), reaches a peak, drops beyond this turning point 
 ($l_{\rm p}\uparrow\Delta\Omega_{\rm p}\downarrow$) and turns to attractive. 

This non-monotonic behaviour indicates that even at large $\mbox{Spd}^{3+}$ densities, the polymer has to overcome an electrostatic barrier at the pore entrance before penetrating the pore by following the downhill grand potential landscape. The presence of the barrier can be explained by noting that for $\kappa_{\rm b} l_{\rm p} \lesssim1$, the self-energy of 
Eq.~(\ref{46}) scales quadratically with the polymer length $l_{\rm p}$. Thus, at the pore entrance, the attractive self-energy is dominated by the repulsive MF-component of Eq.~(\ref{27}) scaling linearly with the length $l_{\rm p}$. 
In Fig~\ref{fig6}(b), we plot the critical penetration length $l_{\rm p}^*$ where the grand potential switches from repulsive to attractive. One notes that the length $l_{\rm p}^*$ drops with increasing $\mbox{Spd}^{3+}$ concentration $\rho_{{\rm b}3+}\uparrow l_{\rm p}^*\downarrow$ and membrane charge $\sigma_{\rm m} \uparrow l_{\rm p}^*\downarrow$. The predictions in this phase diagram call for
verification by translocation experiments. We finally note that the grand potential landscape obtained from Eqs.~(\ref{27}) and~(\ref{46}) can be used to account for electrostatic pore-polymer interactions in the MD simulations~\cite{sim1,sim4,sim5}.

\section{Summary and Conclusions}

One of the most important issues in translocation experiments for biological polyelectrolytes concerns the issue of electrostatic
barriers stemming from the interplay of electrostatic interactions in the system. 
In the present work we have characterized electrostatic polymer-pore interactions in multivalent electrolyte mixtures where 
MF approaches break
down.  We have developed a beyond-MF theory where charge correlations are taken into account by the kernel Eq.~(\ref{15}) that cannot however be exactly solved in a closed form. Instead, we have solved this equation analytically within the WKB approximation. This is the main technical achievement of our work. Our main results and conclusions are summarised below.

The cation attraction into the negatively charged nanopore enhances the screening ability of the pore with respect to the reservoir. This translates into an attractive force that opposes the MF level like-charge repulsion and the repulsive image-charge barrier acting on the polymer. In the case of polymers with charge density above the critical value $\tau_{\rm c}=2/(m\ell_{\rm B})$, upon addition of multivalent cations into the solution, the attractive force takes over the repulsive components and triggers the attraction of the polymer by the like-charged pore. This is the key finding of our work. The cation-induced like-charge attraction mechanism presents itself as an efficient way to enhance the rate of anionic polymer capture by negatively charged Si-based nanopores.  

We found that the minimum multivalent counterion concentration $\rho_{{\rm b}m+}$ for the occurrence of polymer-pore attraction obeys a non-trivial scaling law given by Eq.~(\ref{63})
%
%
which predicts the reduction of the critical cation concentration with the enhancement of the membrane charge density $\sigma_{\rm m}\uparrow\;\rho^*_{{\rm b}m+}\downarrow$ or the reduction of the monovalent salt concentration 
$\rho_{{\rm b}+}\downarrow\;\rho^*_{{\rm b}m+}\downarrow$. These characteristics may provide an accurate control over polymer-pore interactions through the alteration of the membrane charge or salt density. 

Furthermore, we have also scrutinized the effect of pore confinement. We found that in strongly charged pores, the reduction of the radius below the critical value $d^*$ given by Eq.~(\ref{64})
%
%
turns polymer-pore interactions from repulsive to attractive. Interestingly, the radius $d^*$ corresponds to an upper bound for attractive interactions. This implies that at strong enough membrane charge, despite the presence of the image-charge barrier on the polymer, confinement favours the like-charge polymer-pore attraction. These predictions together with the scaling laws of Eqs.~(\ref{63}) and~(\ref{64}) can be beneficial to translocation experiments. Moreover,  our formalism presents itself as a consistent tool to incorporate electrostatic polymer-pore interactions into previous MD simulation algorithms.

Our model is based on some approximations. In order to formulate the problem analytically, we exploited the cylindrical symmetry and neglected edge effects associated with the finite membrane thickness. We believe that this complication can be included exclusively by solving the kernel Eq.~(\ref{9}) numerically on a discrete lattice. Furthermore, our electrostatic formalism is based on the one-loop-level test-charge theory~\cite{Buyuk2017}. This formalism does not cover the electrostatic strong-coupling regime and treats the polymer charges as a perturbation. These limitations can be overcome in a future work by using the variational approach from Hatlo and Lue that can cover charge correlations from weak to strong-coupling regime~\cite{SCHatlo}. Then, for the sake of analytical simplicity, we treated the polymer as a line charge. The standard way to consider the lateral structure of polymers consists in modelling them as rigid cylinders. This extension will require (i) the evaluation of the self-energy of Eq.~(\ref{13}) with the numerical solution of Eqs.~(\ref{37V}) and~(\ref{40}) for finite Fourier modes $n$ and (ii) the inclusion of van der Waals forces resulting from the dielectric contrast between the polymer, the membrane, and the solvent~\cite{PRE2016}. Finally, we have considered here polymer-pore interactions from a purely electrostatic perspective. It should be noted that hydrodynamics of the solvent also plays an important role in polymer capture and translocation~\cite{the7}. Within the framework of our recently developed non-equilibrium polymer translocation model~\cite{mf}, we plan to combine the present electrostatic formalism with hydrodynamic effects in an upcoming work. 

\appendix

\section{Numerical evaluation of the electrostatic Green's function}
\label{apx1}

In this Appendix, we explain the numerical calculation of the Fourier-transformed Green's function $\tv_{n}(r,r';k)$ solving 
Eq.~(\ref{15}). This equation will be solved by iteration around the Donnan Green's function solution to the differential equation
\bea\label{ap1}
&&\left\{\frac{1}{r}\partial_rr\e(r)\partial_r-\e(r)\left[\frac{n^2}{r^2}+k^2+\kappa_{\rm D}^2(r)\right]\right\}\tv_{{\rm D},n}(r,r';k)\nonumber\\
&&=-\frac{e^2}{k_{\rm B}T}\frac{1}{r}\delta(r-r'),
\eea
where we defined the piecewise screening parameter $\kappa_{\rm D}(r)=\kappa_{\rm D}\theta(d-r)$ with 
$\kappa_{\rm D}$ given by Eq.~(\ref{25}). Now, we use the definition of the Green's function
\be\label{ap2}
\int\mathrm{d}\br''v_{\rm D}^{-1}(\br,\br'')v_{\rm D}(\br'',\br')=\delta(\br-\br').
\ee
Inserting the Fourier expansion of Eq.~(\ref{11}) into Eq.~(\ref{ap2}), the latter takes form
\be\label{ap3}
\int_0^{\infty}\mathrm{d}r''r''\tv_{{\rm D},n}^{-1}(r,r'';k)\tv_{{\rm D},n}(r'',r';k)=\frac{1}{r}\delta(r-r').
\ee
By using Eq.~(\ref{ap3}), one can show that the kernel operator associated with Eq.~(\ref{ap1}) is
\bea\label{ap4}
\tv^{-1}_{{\rm D},n}(r,r';k)&=&-\frac{k_{\rm B}T}{e^2}\left\{\frac{1}{r}\partial_rr\e(r)\partial_r\right.\\
&&\left.-\e(r)\left[\frac{n^2}{r^2}+k^2+\kappa_{\rm D}^2(r)\right]\right\}\frac{\delta(r-r')}{r}.\nonumber
\eea
In terms of the operator of Eq.~(\ref{ap4}), one can now express the kernel Eq.~(\ref{15}) as
\bea\label{ap5}
&&\int_0^\infty\mathrm{d}r_1r_1\tv_{{\rm D},n}^{-1}(r'',r_1;k)\tv_{n}(r_1,r';k)\nonumber\\
&&=\frac{1}{r''}\delta(r''-r')+\delta n(r'')\tv_{n}(r'',r';k),
\eea
where we defined the local screening correction
\be\label{ap6}
\delta n(r)=\sum_{i=1}^p\rho_{{\rm b}i}q_i^2\left[e^{-q_i\phi_{\rm D}}-e^{-q_i\phi_{\rm m}(r)}\right]\theta(d-r).
\ee
In Eq.~(\ref{ap6}), the pore potential $\phi_{\rm m}(r)$ corresponds to the exact numerical solution of the PB Eq.~(\ref{14}). 
Multiplying now Eq.~(\ref{ap5}) by $r''\tv_{{\rm D},n}(r,r'';k)$, integrating over the variable $r''$, and using 
Eq.~(\ref{ap3}), Eq.~(\ref{15}) can be finally converted to the following integral relation
\bea
\label{ap7}
\tv_{n}(r,r';k)&=&\tv_{{\rm D},n}(r,r';k)\\
&&+\int_0^\infty\mathrm{d}r''r''\tv_{{\rm D},n}(r,r'';k)\delta n(r'')\nonumber\\
&&\hspace{1.6cm}\times\tv_{n}(r'',r';k).\nonumber
\eea

The iterative solution of Eq.~(\ref{ap7}) requires the knowledge of the Donnan Green's function $\tv_{{\rm D},n}(r,r';k)$. 
In the present case where ions are located in the nanopore, i.e. $r<d$ and $r'<d$, the solution to Eq.~(\ref{ap1}) satisfying the boundary conditions Eqs.~(\ref{19}) - (\ref{22}) reads~\cite{Buyuk2014}
\bea\label{ap8}
\tv_{{\rm D},n}(r,r';k)&=&4\pi\ell_{\rm B}\left[\mathrm{K}_n(p_{\rm D}r_>)\mathrm{I}_n(p_{\rm D}r_<)\right.\\
&&\hspace{9mm}\left.+F_n(k)\mathrm{I}_n(p_{\rm D}r_<)\mathrm{I}_n(p_{\rm D}r_>)\right].\nonumber
\eea
In Eq.~(\ref{ap8}), we used the radial variables of Eq.~(\ref{41}), and introduced the parameter $p_{\rm D}=\sqrt{\kappa_D^2+k^2}$ and the auxiliary function accounting for the dielectric nanopore 
\be\label{ap9}
F_n(k)=\frac{p_{\rm D} \mathrm{K}_n\left(|k|d\right)\mathrm{K}'_n(p_{\rm D} d)-\gamma |k| \mathrm{K}_n(p_{\rm D} d)\mathrm{K}'_n\left(|k| d\right)}{\gamma |k| \mathrm{I}_n(p_{\rm D} d)\mathrm{K}'_n\left(|k| d\right)-p_{\rm D} \mathrm{K}_n\left(|k| d\right)\mathrm{I}'_n(p _{\rm D}d)}
\ee
with $\gamma=\e_{\rm m}/\e_{\rm w}$. In order to solve Eq.~(\ref{ap7}) by iteration, at the first iterative step, we solve numerically the PB Eq.~(\ref{14}) and calculate the radial integral in Eq.~(\ref{ap7}) by replacing the Green's function $\tv_{n}(r,r';k)$ by the Donnan propagator of Eq.~(\ref{ap8}). The output propagator is injected into the integral at the next iterative step and this cycle is continued until numerical convergence is achieved. We also note that in the thermodynamic limit $l_{\rm p}\to\infty$ where the infrared limit $k\to0$ of the Green's function solely contributes to the polymer self-energy, the auxiliary function of Eq.~(\ref{ap9}) takes the simpler form
\be\label{ap10}
F_n(0)=\frac{\kappa_{\rm D}d\;\mathrm{K}_{|n|-1}(\kappa_{\rm D} d)+(1-\gamma)|n|\mathrm{K}_n(\kappa_{\rm D} d)}{\kappa_{\rm D}d\;\mathrm{I}_{|n|-1}(\kappa_{\rm D} d)-(1-\gamma)|n|\mathrm{I}_n(\kappa_{\rm D} d)}.
\ee


\begin{thebibliography}{99}
\bibitem {not2} This charge density value was determined in Ref.~\cite{the7} by fitting the experimental charge conductivity data of nanopores blocked by a double-stranded DNA molecule with radius $a\approx1$ nm.
\end{thebibliography}
\end{document}